\def\lam{\lambda}
\def\kap{\kappa}
\def\alam{A_\lambda}
\def\akap{A_\kappa}

\def\tanb{\tan\beta}

\def\br{{\rm Br}}

\def\anti{\overline}

\def\ma{m_a}
\def\mh{m_h}
\def\hi{h_1}

\def\ai{a_1}
\def\mhi{m_{h_1}}

\def\mai{m_{a_1}}
\def\mueff{\mu_{\rm eff}}
\def\mtau{m_\tau}

\def\calo{{\cal O}}

\def\mb{m_b}
\def\mz{m_Z}

\def\lsim{\mathrel{\raise.3ex\hbox{$<$\kern-.75em\lower1ex\hbox{$\sim$}}}}
\def\gsim{\mathrel{\raise.3ex\hbox{$>$\kern-.75em\lower1ex\hbox{$\sim$}}}}

\def\gev{~{\rm GeV}}

\def\noi{\noindent}

\def\mhusq{m_{H_u}^2}
\def\mhdsq{m_{H_d}^2}
\def\mssq{m_S^2}

\def\vev#1{\langle #1 \rangle}

\def\cbb{C_{eff}^{2b}}

\def\tauptaum{\tau^+\tau^-}
\def\hsm{h_{\rm SM}}

\newcommand{\nc}{\newcommand}
\nc{\beq}{\begin{equation}}   \nc{\eeq}{\end{equation}}
\nc{\bea}{\begin{eqnarray}}   \nc{\eea}{\end{eqnarray}}
\nc{\baa}{\begin{array}}      \nc{\eaa}{\end{array}}
\nc{\bit}{\begin{itemize}}    \nc{\eit}{\end{itemize}}
\nc{\ben}{\begin{enumerate}}  \nc{\een}{\end{enumerate}}
\nc{\bce}{\begin{center}}     \nc{\ece}{\end{center}}
\def\beqa{\begin{eqnarray}}
\def\eeqa{\end{eqnarray}}
\def\bed{\begin{description}}
\def\eed{\end{description}}

\documentclass[
    ,final            
  ]
  {aipproc}

\layoutstyle{6x9}

\begin{document}

\title{Naturalness of EWSB in NMSSM \\ and Higgs at~100~GeV}


\author{Radovan Derm\' \i\v sek}{
  address={School of Natural Sciences, Institute for Advanced
Study, Princeton, NJ 08540} }

\author{John F. Gunion}{
  address={Department of Physics, University of California at
Davis, Davis, CA 95616} }

\begin{abstract}
Naturalness of electroweak symmetry breaking in 
the next-to-minimal supersymmetric model 
points to scenarios in which the lightest CP-even Higgs boson has SM-like $ZZh$ coupling and mass 
$m_h\sim 100 $ GeV   and decays partly via
$h\to b\bar b$  but dominantly into two CP-odd 
Higgs bosons, $h \to aa$, where $m_a<2m_b$ so that $a\to \tau^+ \tau^-$
(or light quarks and gluons) decays are dominant. 
These preferred scenarios correlate well with the observed excess of events at LEP for $m_h\sim 100 $ GeV.

\end{abstract}

\maketitle



In the minimal supersymmetric standard model (MSSM) a generic supersymmetric (SUSY) 
spectrum 
that leads to natural electroweak symmetry breaking predicts a mass for the Higgs boson 
which is already ruled out by LEP data. To satisfy LEP limits the SUSY spectrum 
has to be either very heavy or very special.\footnote{Scenarios that lead to such a special
SUSY spectrum were recently found, see for example mixed
anomaly-modulus mediation~\cite{modulus_anomaly} or gauge
mediation with gauge messengers~\cite{gauge_messengers}. Both
scenarios generate large mixing in the stop sector which maximizes
the Higgs mass allowing all experimental limits to be satisfied with
a fairly light SUSY spectrum.}
However in the next-to-minimal supersymmetric model (NMSSM) the Higgs boson 
can be where it is predicted from natural EWSB with a generic SUSY 
spectrum~\cite{Dermisek:2005ar} and 
there is even a hint in LEP data that this might be the case~\cite{Dermisek:2005gg}.

The NMSSM is a very attractive model.
It provides an elegant solution to the $\mu$ problem of the
MSSM via the introduction of a singlet superfield $\widehat{S}$.  For
the simplest possible scale invariant form of the superpotential, 
\vspace*{-.09in} \beq \label{1.1} \lambda \ \widehat{S}
\widehat{H}_u \widehat{H}_d + \frac{\kappa}{3} \ \widehat{S}^3,
\vspace*{-.11in} \eeq \noi
the
scalar component of $\widehat{S}$ naturally acquires a vacuum
expectation value of the order of the SUSY breaking scale, giving
rise to a value of $\mu$ of order the electroweak scale. 
Its particle content differs from the
MSSM by the addition of one CP-even and one CP-odd state in the
neutral Higgs sector (assuming CP conservation), and one
additional neutralino.  We follow the conventions of
\cite{Ellwanger:2004xm}. 
The associated trilinear soft terms are
\vspace*{-.1in}
\beq \label{1.2}
\lambda A_{\lambda} S H_u H_d + \frac{\kappa} {3} A_\kappa S^3 \,.
\vspace*{-.1in}
\eeq
%
The final two input parameters are
%
$\tan \beta = h_u/h_d$ and $\mu_\mathrm{eff} = \lambda
s $,
%
where $h_u\equiv
\vev {H_u}$, $h_d\equiv \vev{H_d}$ and $s\equiv \vev S$.
These, along with $\mz$, can be viewed as
determining the three SUSY breaking masses squared for $H_u$, $H_d$
and $S$ (denoted $\mhusq$, $\mhdsq$ and $\mssq$)
through the three minimization equations of the scalar potential.
Thus,  the
Higgs sector of the NMSSM is described by the six parameters
$\lambda\ , \ \kappa\ , \ A_{\lambda} \ , \ A_{\kappa}, \ \tan \beta\ ,
\ \mu_\mathrm{eff}$.


\begin{figure}[t]
\hspace{-0.5cm}  
\includegraphics[scale=0.30,angle=90,clip=true]{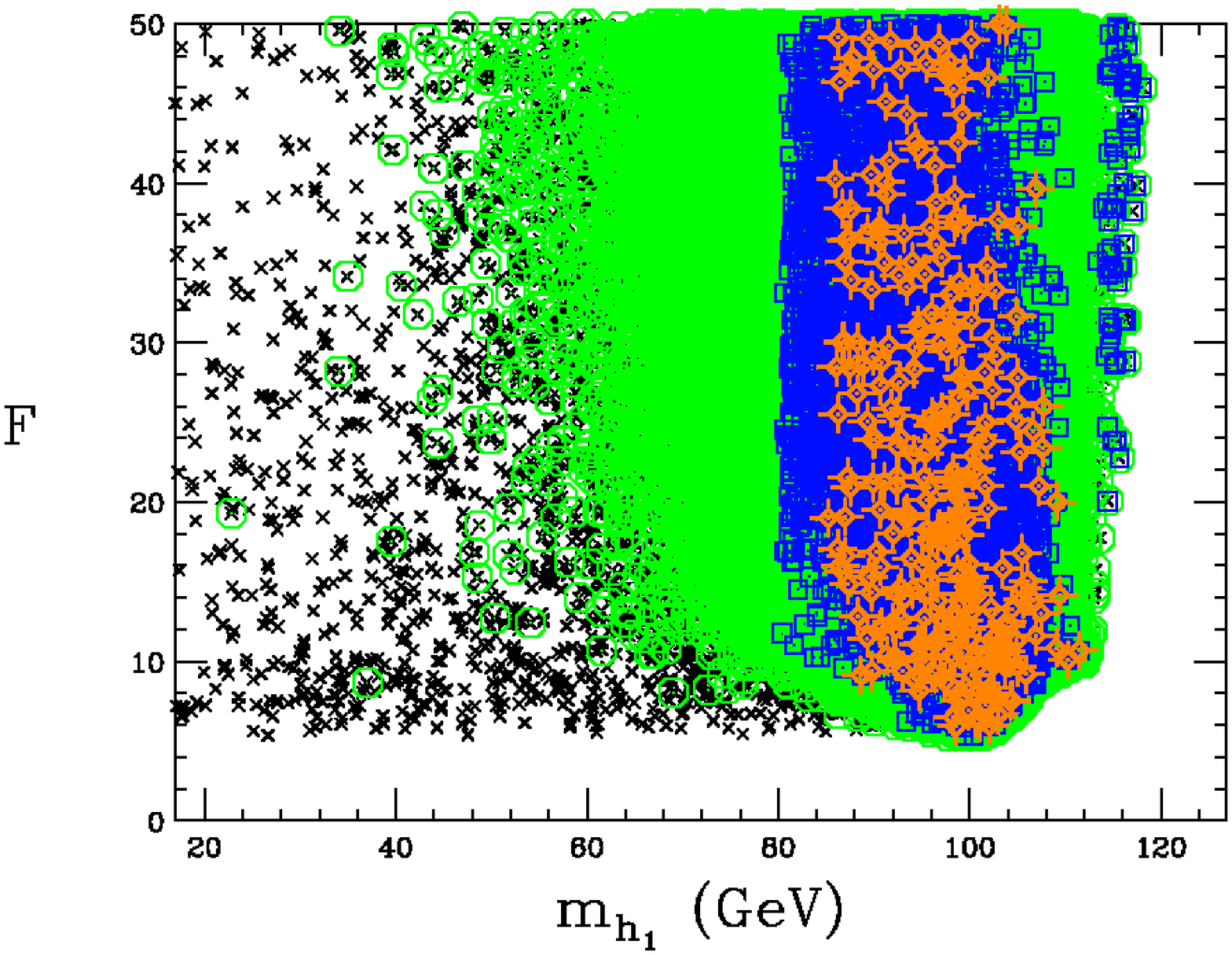}
\hspace{0.5cm}
\includegraphics[scale=0.28,angle=90,clip=true]
{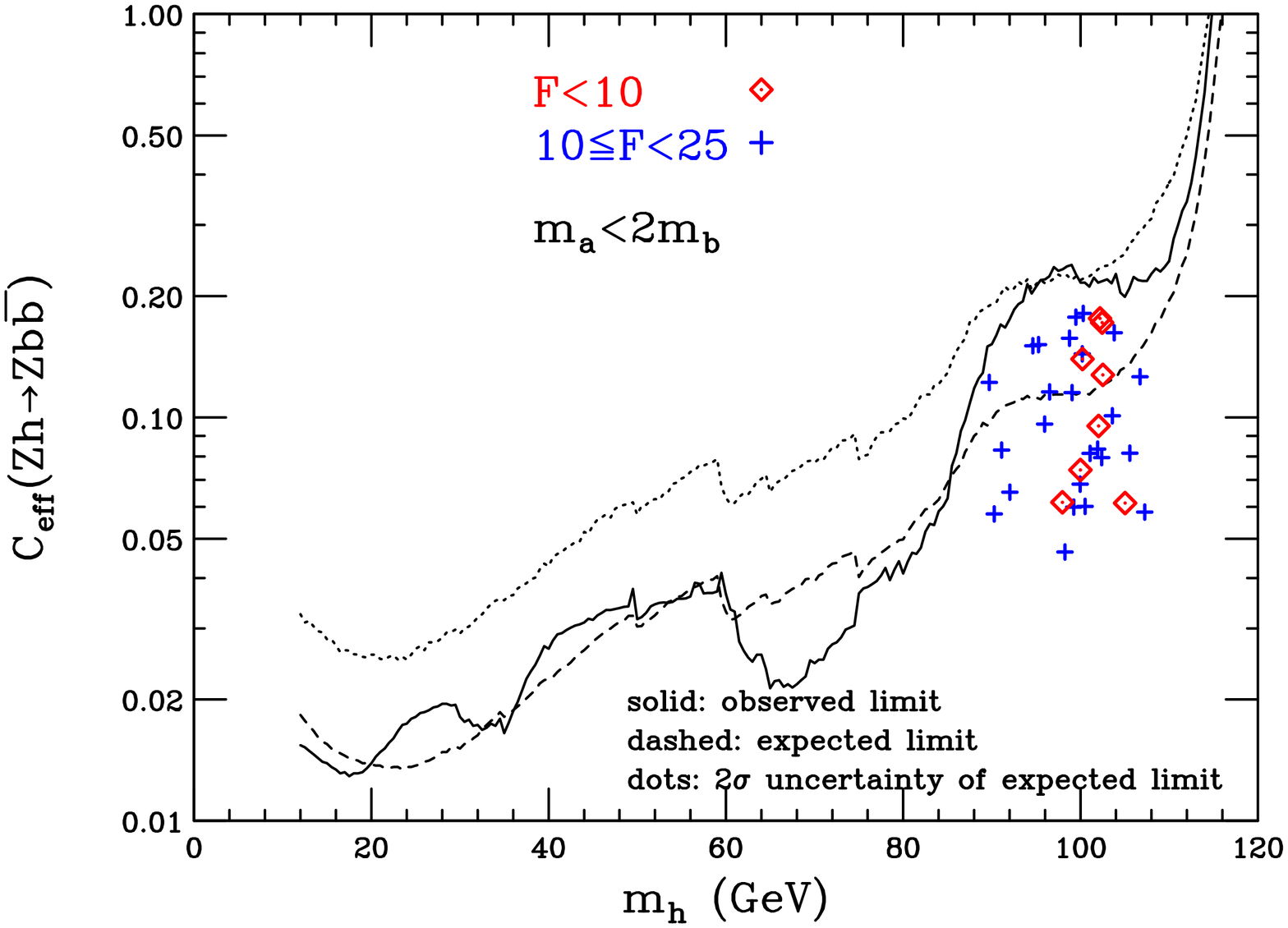}
  \caption{
  Left: fine tuning vs. the Higgs mass for randomly generated NMSSM parameter choices 
  with $\tanb=10$ and  $M_{3}(\mz)=300\gev$. Green circles are the
  points that satisfy limits on SUSY spectrum. 
  Blue squares either have $\mhi > 114.4$ GeV or, if $\mhi<114.4$ GeV,
  large branching fraction for $\hi \to \ai \ai$.
  Orange stars represent scenarios in which $\hi \to \ai \ai$ is large and $\mai < 2 m_b$.
  Right:
expected and observed 95\% CL limits on
  $\cbb=[g_{ZZh}^2/g_{ZZ\hsm}^2]\br(h\to b\anti b)$  are shown vs. $\mh$.  
  Also plotted are the
  predictions for NMSSM scenarios that appeared  in our 
  original (low density)  scan~\cite{Dermisek:2005ar} 
  that give
  fine-tuning measure $F<25$ and $\mai<2\mb$ and that are consistent
  with the preliminary LHWG analysis code.
\label{finetuning}
}
\end{figure}

We can use naturalness of EWSB to predict the mass of the Higgs boson. 
Let us  define the measure of fine tuning as
\vspace*{-.1in}
\beq F={\rm Max}_p F_p \equiv {\rm
  Max}_P\left|{d\log \mz\over d\log p}\right|\,,
\vspace*{-.1in}  
\eeq
where the parameters $p$ comprise the GUT-scale soft-SUSY-breaking
parameters.  We randomly generate NMSSM scenarios with fixed $\tan \beta = 10$ and $M_3(\mz) = 300$ GeV and plot fine tuning vs. the Higgs mass, see Fig.~\ref{finetuning} (left). Scenarios that satisfy the experimental constraints on superpartner masses are labeled by green circles.
Scenarios labeled by  blue squares have in addition $\mhi > 114.4$ GeV
and/or  large ${\rm Br}(\hi \to \ai \ai)$ (typically, the lightest CP-odd Higgs boson is mostly singlet). If $\mai > 2m_b$ the later scenarios are further constrained by $Zh\to Zaa\to Zb\anti b b\anti b$ limits and those with full strength $ZZh$ coupling are ruled out unless $\mhi \gsim 110$ GeV. 
However, there are no limits on the $Zh\to
Zaa\to Z\tauptaum\tauptaum$ final state for $\mh\gsim 87\gev$ and so the scenarios with $\mai < 2m_b$
satisfy all experimental constraints. These scenarios (orange stars in Fig.~\ref{finetuning}) are among the scenarios with the very lowest fine tuning one can find given the current constraints on superpartners. 
Naturalness of EWSB in NMSSM clearly points to  
scenarios in which the Higgs boson with SM-like $ZZh$ coupling and mass 
$\mhi \sim 100$ GeV
 decays partly via
$\hi\to b\bar b$  but dominantly into two CP-odd 
Higgs bosons, $\hi \to \ai\ai$, where $\mai<2m_b$ so that $\ai \to \tau^+ \tau^-$
(or light quarks and gluons) decays are dominant. 
Further, since the $ZZh$ and $WWh$ couplings are
very SM-like, these scenarios give excellent agreement with precision
electroweak data.

Quite interestingly, the largest excess of events in the search for the SM Higgs boson at LEP is 
for a test Higgs mass of $\mh\sim 100\gev$~\cite{Barate:2003sz}. This
excess cannot be interpreted as the SM Higgs boson (a SM Higgs boson
with a mass of $100$ GeV would lead to $\sim 10$ times larger excess). However, in the orange scenarios discussed above, the Higgs boson has
SM-like $ZZh$ coupling, but highly
reduced $\br(h\to b\anti b)$; such scenarios can naturally explain the observed excess. 
In Fig.~\ref{finetuning} (right) we plot  $\cbb$ limits and $\cbb$ predictions for NMSSM scenarios that appeared  in our 
  original (low density)  scan~\cite{Dermisek:2005ar} 
  that give
  fine-tuning measure $F<25$ and $\mai<2\mb$.
  All these points 
are consistent with the experimental and theoretical constraints
built into NMHDECAY~\cite{Ellwanger:2004xm} as well as with limits from the preliminary LHWG
full confidence level/likelihood analysis (we thank P. Bechtle for doing this analysis).  
The eight points with the lowest fine tuning, $F<10$,
cluster near $\mhi\sim 98\div105\gev$ (see also Fig.~3
of \cite{Dermisek:2005ar}). 
A detailed description of these eight points, 
including precise masses and branching
ratios of the $\hi$ and $\ai$,  can be found in Ref.~\cite{Dermisek:2005gg}.
A significant
fraction of the $F<10$ points with low fine tuning are very consistent with the observed
event excess.

Finally, let us discuss conditions under which scenarios with $\mai < 2m_b$ and large Br($\hi\to \ai\ai$) 
occur in the NMSSM.
In the NMSSM, one of the CP-odd Higgses is massless in the
Peccei-Quinn symmetry limit, $\kappa \to 0$, or in the R-symmetry
limit, $A_\kappa, A_\lambda \to 0$~\cite{Dobrescu:2000jt}. 
The masslessness of $a_1$ in the limit $A_\kappa, A_\lambda \to 0$
can be understood as a consequence of a global $U(1)_R$ symmetry
of the superpotential under which the charge of $S$ is half of the
charge of $H_u H_d$. In the limit $A_\kappa, A_\lambda \to 0$ it
is also a symmetry of the scalar potential. This symmetry is
spontaneously broken by the vevs of $H_u$, $H_d$ and $S$,
resulting in a (massless) Nambu-Goldstone boson in the spectrum. Soft
trilinear couplings explicitly break $U(1)_R$ and thus lift the
mass of the $a_1$. For small trilinear couplings, the mass of the
lightest CP-odd Higgs boson is approximately given as:
\vspace*{-.1in}
\begin{equation}
m_{a_1}^2  \simeq 3s \left( \frac{3 \lambda A_\lambda \cos^2
\theta_A }{2 \sin 2 \beta}
 - \kappa  A_\kappa \sin^2 \theta_A \right), \label{eq:ma1}
\vspace*{-.1in}
\end{equation}
where $\cos \theta_A$ measures the doublet component of the
lightest CP-odd Higgs mass eigenstate,
$a_1 =  \cos \theta_A \, A_{MSSM} + \sin \theta_A \, A_S$.
In the limit of large $\tan \beta$ or $|s| \gg v$, $\cos\theta_A$ can be
approximated by
$\cos \theta_A  \simeq 2v/s \tan \beta$,
which indicates that
in
these limits the lightest mass eigenstate is mostly singlet.

Naively, an arbitrarily small mass for the $\ai$ is achievable provided
small values of $A_\kappa$ and $A_\lambda$ are generated by a SUSY
breaking scenario. 
However, 
$A_\kappa$ and $A_\lambda$ 
receive radiative corrections from gaugino masses
($A_\kappa$ only at the two-loop level) and we should
naturally expect
$A_\lambda (\mz) \simeq {\cal O}(M_2(\mz)) \gsim 100\gev$ and $A_\kappa(\mz) \gsim few\gev$.
Much smaller values would require cancellations between the values
of $A_\lambda$, $A_\kappa$ coming from a particular SUSY breaking
scenario and the contributions from the radiative corrections.

\begin{figure}

\includegraphics[scale=0.25,angle=90,clip=true]
{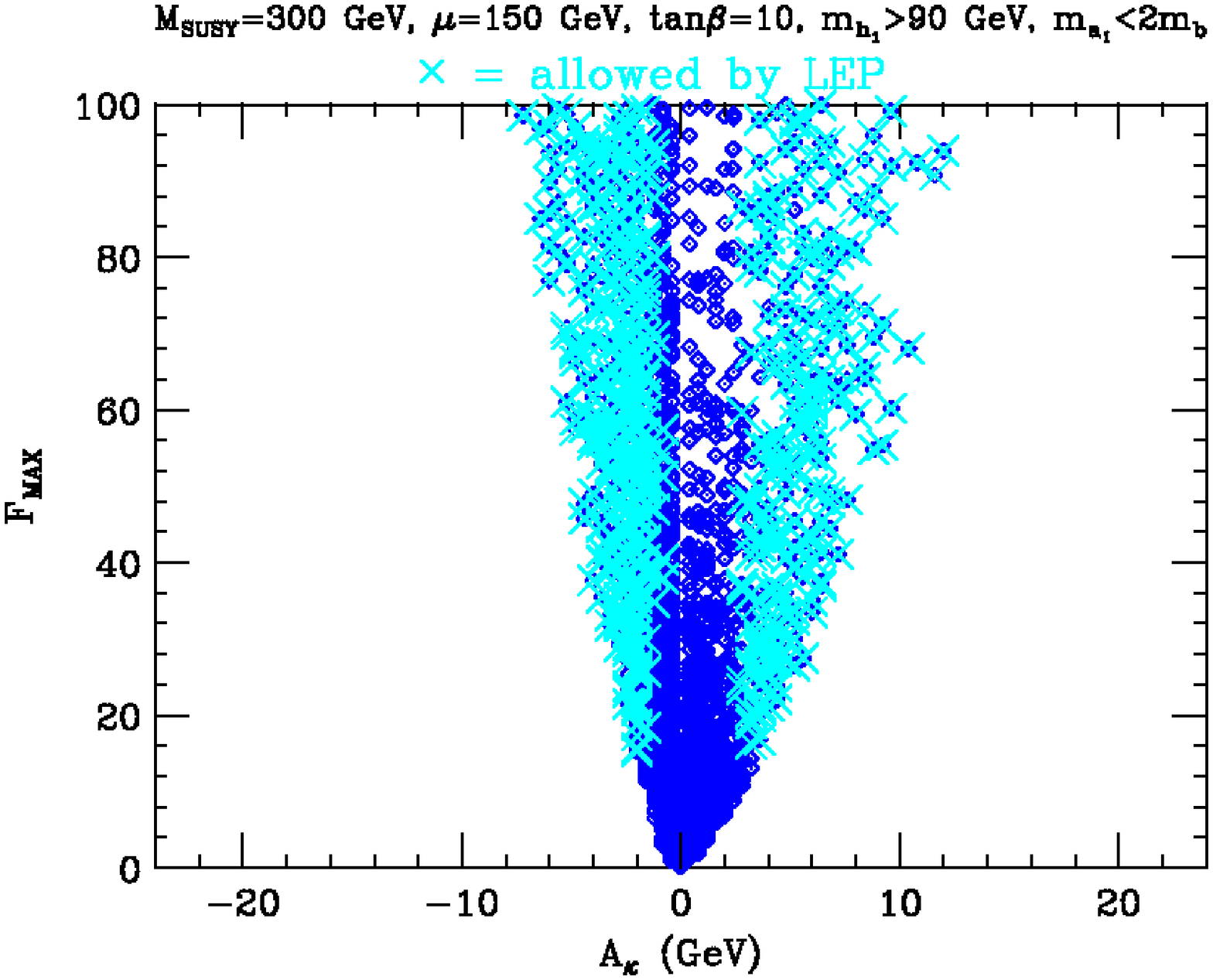}
\hspace{1.cm}
\includegraphics[scale=0.25,angle=90,clip=true]
 {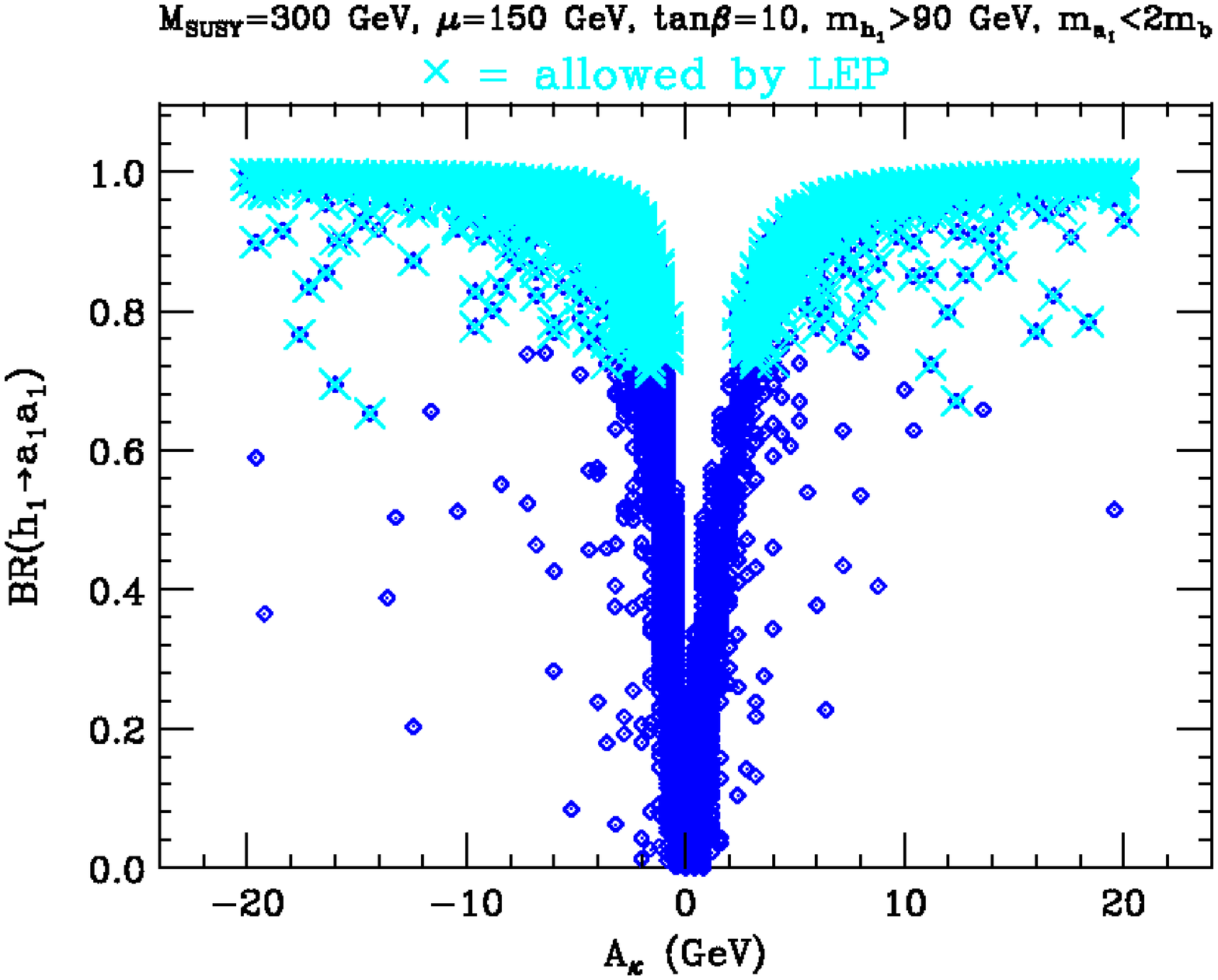}
  \caption{Tuning in $m_{a_1}^2$ vs. $A_\kappa$ (left)
 and Br($\hi\to\ai\ai$) vs. $A_\kappa$ (right)
  for $\mu=150\gev$ and $\tanb=10$.
Light grey (cyan) large crosses are
  points that satisfy all
experimental limits. The dark (blue) diamonds are those points that
do not have large enough $\br(\hi\to\ai\ai)$ so as to suppress $\br(\hi\to
b\bar b)$ sufficiently to escape LEP limits.}
\label{fig:fmax_and_Br}
\end{figure}

Eq.~(\ref{eq:ma1}) indicates that  
the contribution from $A_\lambda\gsim 100$ GeV to $\mai^2$ is highly
suppressed if the lightest CP-odd Higgs is mostly singlet. 
We define a  measure of the
tuning in $A_\lambda(\mz)$ and $A_\kappa(\mz)$ necessary to achieve $\mai <2 m_b$ as
$F_{MAX} = \max \left\{ \left| d \log \mai^2 / d \log A_\lambda(\mz) \right|,
\left| d \log \mai^2 / d \log A_\kappa(\mz) \right| \right\}$
%
which is to be evaluated for given choices of $\alam(\mz)$ and
$\akap(\mz)$ that yield a given $\mai$~\cite{DG-new}.  This definition
reflects the fact that $\mai$ is completely determined by these
parameters for fixed $\lam,\kap,\mueff,\tanb$.  
The dependence of
$F_{MAX}$ on $A_\kappa$ and $A_\lambda$ is given in
Fig.~\ref{fig:fmax_and_Br}. 
As expected, the smallest fine
tuning or sensitivity is achieved for as small $A_\kappa$ and
$A_\lambda$ as possible. Of course, as we discussed earlier, very
small values of $A_\kappa$ and $A_\lambda$ would require
cancellations between the bare values and the RGE-induced
radiative corrections and this kind of cancellation would not be
visible from the definition of $F_{MAX}$. However, this is not a a
concern given that very small values of $A_\kappa$
and $A_\lambda$ do not in any case lead to sufficiently large 
Br$(\hi\to\ai\ai)$ that $\br(\hi\to b\anti b)$ is adequately
suppressed, see Fig.~\ref{fig:fmax_and_Br}. Thus, in the region of parameter space where soft
trilinear couplings are at least of order the typical RGE-induced
contributions, which is also the region where $\br(\hi\to\ai\ai)$
is large, the tuning of the soft trilinear couplings can be as small
as $\calo(5\% - 10\%)$. 
Finally, we would like to note that $F_{MAX}$, 
typically substantially overestimate the magnitude of fine
tuning with respect to GUT scale parameters.  For scenarios in which gaugino masses 
and soft-trilinear couplings are correlated (not free parameters but
rather calculable from a SUSY breaking scale) there is in principle no
tuning 
associated with the  SUSY breaking necessary to achieve 
$\mai < 2 m_b$~\cite{DG-new}.

In conclusion, we reemphasize that in the NMSSM the SM-like Higgs boson can be where it is 
 predicted on the basis of natural EWSB and that 
the LEP event excess
in the $Z+b$'s channel for a reconstructed Higgs mass of $\mh\sim 100\gev$
is consistent with a scenario in which
the $h$
decays mainly via $h\to aa\to \tauptaum\tauptaum$ ($2\mtau<\ma<2\mb$)
or $4~jets$ ($\ma<2\mtau$)
leaving an appropriately reduced rate for $h\to b\anti b$.
We speculate that similar
results will emerge in other supersymmetric models with a Higgs sector
that is more complicated than that of the CP-conserving MSSM.
This suggests that the strategy for Higgs searches at the Tevatron and LHC should be modified.


\vspace{0.3cm}
RD thanks K.~Agashe, P.~Schuster, M.~Strassler and N.~Toro for discussions during 
the conference. RD
is supported by DOE grant
DE-FG02-90ER40542. JFG is supported by DOE grant \#DE-FG02-91ER40674 and by the
U.C. Davis HEFTI program.


\vspace{-.2cm}
\begin{thebibliography}{99}


\bibitem{modulus_anomaly}
  R.~Kitano, these proceedings; Y.~Nomura, these proceedings.

\bibitem{gauge_messengers}
  R.~Dermisek and H.~D.~Kim,
  Phys.\ Rev.\ Lett.\  {\bf 96}, 211803 (2006)
  [arXiv:hep-ph/0601036];
%
  R.~Dermisek, H.~D.~Kim and I.~W.~Kim,
  arXiv:hep-ph/0607169.


\bibitem{Dermisek:2005ar}
  R.~Dermisek and J.~F.~Gunion,
  Phys.\ Rev.\ Lett.\  {\bf 95}, 041801 (2005)
  [arXiv:hep-ph/0502105].


\bibitem{Dermisek:2005gg}
  R.~Dermisek and J.~F.~Gunion,
  Phys.\ Rev.\ D {\bf 73}, 111701 (2006)
  [arXiv:hep-ph/0510322].
  
\bibitem{Ellwanger:2004xm}
U.~Ellwanger, J.~F.~Gunion and C.~Hugonie,
arXiv:hep-ph/0406215.
  
\bibitem{Barate:2003sz}
  R.~Barate {\it et al.},  
  Phys.\ Lett.\ B {\bf 565}, 61 (2003)
  [arXiv:hep-ex/0306033].
      
\bibitem{Dobrescu:2000jt}
B.~A.~Dobrescu, G.~Landsberg and K.~T.~Matchev,
Phys.\ Rev.\ D {\bf 63}, 075003 (2001).
B.~A.~Dobrescu and K.~T.~Matchev,
JHEP {\bf 0009}, 031 (2000).

\bibitem{DG-new}
    R.~Dermisek and J.~F.~Gunion, 
  arXiv:hep-ph/0611142.

\end{thebibliography}
\end{document}